\documentclass{article}

\usepackage{authblk}
\usepackage{amssymb}
\usepackage{mathtools}
\usepackage[a4paper, portrait, margin=1.5cm]{geometry}
\usepackage{hyperref}
\usepackage{tikz}
\usepackage{braket}

\usepackage{microtype}
\usepackage{ngerman}
\title{Ising formulations of routing optimization problems}

\author[1]{Daniel Jaroszewski}
\author[1]{Fabian Klos}%Fabian Klos \and Benedikt Sturm}
\author[1]{Benedikt Sturm}

\affil[1]{Frankfurt Consulting Engineers GmbH, HOLM, \authorcr \textit{Bessie-Coleman-Str. 7, D-60549 Frankfurt am Main, Germany}}

\date{}

\begin{document}

\twocolumn[{%
  \begin{@twocolumnfalse}
    \maketitle
    \begin{abstract}
    	We formulate binary optimization functions for single-vehicle routing, travelling salesperson and collision-free multi-vehicle routing with significant improvements in the number of variables over existing formulations. The provided functions are readily implemented on gate-based quantum computers using variational algorithms and on adiabatic quantum hardware.
    \end{abstract}
    \vspace{.6cm}
  \end{@twocolumnfalse}
}]

\section{Introduction}

The Quantum Approximate Optimization Algorithm \cite{farhi2014quantum} (QAOA)  inspired by Adiabatic Quantum Computing \cite{farhi2000quantum, farhi2002quantum, crosson2014different} is a  general-purpose algorihm approximating the global optimum of a quadratic binary optimization polynomial
\begin{align*}
	C:\{0,1\}^n&\rightarrow\mathbb{R}_{\geq 0} \\
	\{X_i\}\;\,&\mapsto C(X_1, ..., X_n).
\end{align*}
Its runtime and quality of results on NISQ-era devices \cite{Preskill_2018, farhi2017quantum} crucially depends on the number of binary variables $n$ and the polynomial structure of $C$. It is hence important to phrase a given optimization problem in a suitable optimization function, in particular with a minimum number of binary variables.

In this note, we discuss multiple routing problems in (un)directed weighted graphs $G=(V,A,w)$. Here, $V$ is the set of vertices and $A\subset V \times V$ is the set of weighted arcs each described by a pair $(s,t)\in A$ of source vertex $s$ and target vertex $t$. The weight function $w:A\rightarrow \mathbb{R}_{>0}$ assigns a positive real value to each arc. For each problem in question we formulate an optimization polynomial $C$ whose global minimum corresponds to an optimal solution.

We first discuss single-vehicle routing in chapter \ref{sec:single vehicle routing}. As visualized in figure \ref{fig:single-vehicle routing}, the goal is to find a path from a specified origin $o \in V$ to a destination $d \in V$ such that the sum of weights of all edges on the path is minimal. We provide an optimization function for a directed graph whose number of binary variables $n$ equals the number of edges without relying on any classical preprocessing. This implies a significant advantage over the Quantum version of the A*-algorithm of \cite{bauckhage2018towards}.

In section \ref{sec: time dependent phenomena}, we discuss a different model for single-vehicle routing which allows for the inclusion of time-dependent external phenomena like weather or traffic. This model creates a new graph $G'$ out of $G$. Hence, it requires preprocessing and is particularly suited for scenarios where weights are adjusted, but the overall graph is constant. A natural field of application are therefore aviation industries with similar paths every few minutes but rapidly changing external conditions.

This model paves the way towards the formulation of the travelling salesperson, see section \ref{sec:tsp}. Here, the goal of selecting a single path connecting all vertices while visiting each vertex exactly once is typically achieved by introducing $|V|-1$ time slices, see e.g. \cite{IsingOverview}. By classical preprocessing, we can reduce the number of Qubits from the resulting $n:= (|V|-1)^2$ significantly depending on the connectivity of the graph.

Finally, we discuss collision-free multi-vehicle routing: The goal is to find paths in $G$ for $N$ vehicles with origins $o_1, ..., o_N\in V$ and destinations $d_1, ..., d_N\in V$. At each point in time, which is properly discussed in section \ref{sec:MAC}, no two vehicles may share the same vertex. The mathematical model is based on the time-dependent model for single-vehicle routing and therefore naturally allows for time-dependent external phenomena. This model is particularly suited for multi-aircraft-control.

In the appendix we provide short overviews of general-purpose optimization with QAOA on gate-model quantum computers and adiabatic quantum optimization.

Finally, note that all models discussed also apply to unweighted and undirected graphs by choosing $w$ constant or doubling the number of edges, respectively.

\section{Single-vehicle routing}
\label{sec:single vehicle routing}

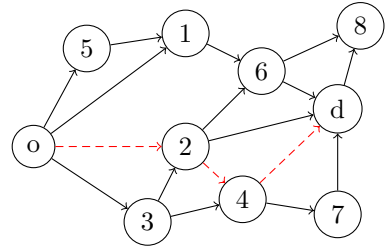
\begin{figure}
	\centering
	\begin{tikzpicture}
		\node[shape=circle,draw=black] (o) at (0,0) {o};
		\node[shape=circle,draw=black] (d) at (4,.5) {d};
		\node[shape=circle,draw=black] (1) at (2,1.5) {1};
		\node[shape=circle,draw=black] (2) at (2,0) {2};
		\node[shape=circle,draw=black] (3) at (1.5,-1) {3};
		\node[shape=circle,draw=black] (4) at (2.75,-.7) {4};
		\node[shape=circle,draw=black] (5) at (.7,1.3) {5};
		\node[shape=circle,draw=black] (6) at (3,1) {6};
		\node[shape=circle,draw=black] (7) at (4,-.9) {7};
		\node[shape=circle,draw=black] (8) at (4.3,1.6) {8};
		\draw[->] (o) -- (1);
		\draw[->] (o) -- (5);
		\draw[->, red, densely dashed] (o) -- (2);
		\draw[->] (o) -- (3);
		\draw[->] (5) -- (1);
		\draw[->] (1) -- (6);
		\draw[->] (6) -- (8);
		\draw[->] (3) -- (2);
		\draw[->] (3) -- (4);
		\draw[->] (4) -- (7);
		\draw[->, red, densely dashed] (4) -- (d);
		\draw[->] (7) -- (d);
		\draw[->] (d) -- (8);
		\draw[->] (6) -- (d);
		\draw[->] (2) -- (d);
		\draw[->, red, densely dashed] (2) -- (4);
		\draw[->] (2) -- (6);
	\end{tikzpicture}
	\caption{An example of single-vehicle routing from origin $o$ to destination $d$. A valid route is a selection of one outgoing arc at $o$, one incoming arc at $d$ and either zero or two arcs at all other vertices.}
	\label{fig:single-vehicle routing}
\end{figure}

A route of length $L$ from origin $o\in V$ to destination $d\in V$ in a directed graph is a collection of arcs
\[\{(s_0, t_0), ..., (s_L, t_L)\} \subset A\] subject to
\begin{align*}
	s_0 &= o \\
	t_k &= s_{k+1} \text{ for all } k=0, ..., L-1 \\
	t_L &= d.
\end{align*}
For the formulation as quadratic unconstrained binary optimization problem, we associate to each arc $(s,t)$ a binary variable $X_{(s,t)}\in\{0,1\}$ which is set to $1$ if the arc is part of the route and set to $0$ otherwise.

The idea behind the optimization objective is the following: Choose arcs of total minimal costs such that origin and destination have one connected arc and all other vertices have either zero or two connected arcs. An example is illustrated in figure \ref{fig:single-vehicle routing}. 

Mathematically, the optimization function to be minimized consists of multiple terms. First, the total costs of the selected arcs needs to be minimized.
\[
	\sum_{(s,t)\in A} w(s,t)\cdot X_{(s,t)}
\]

All other terms enforce constraints such that the selection of arcs corresponds to a valid route. For ease of notation we define the following conditional sum.
\[
	S(\text{cond}) := \sum_{\substack{(s,t)\in A \\ \text{fulfilling condition } \\ \text{cond}}} X_{(s,t)}
\]
For example, the following penalty term in the optimization function contains a sum over all binary variables associated to arcs with source $o$.
\[
	P\cdot\left(S(s=o) -1\right)^2
\]
For $P\in\mathbb{R}_+$ big enough, it ensures that exactly one outgoing arc at the origin is selected. With the following term we additionally exclude all incoming arcs at the origin.
\[
	P\cdot S(t=o).
\]
Generally, the penalty $P\in\mathbb{R}_+$ must be greater than the sum of all arc weights. Similar terms ensure one incoming, but no outgoing arc at the destination vertex.

Each vertex different from origin and destination must either have zero or one outgoing arc
\[
	P\sum_{v\in V\setminus\{o,d\}} \left(S(s=v) -1\right) \cdot S(s=v).
\]
A similar term with $s$ replaced by $t$ ensures zero or one incoming arcs at each central vertex. Finally, at all those vertices the number of incoming and outgoing arcs must equal:
\[
	P\sum_{v\in V\setminus\{o,d\}} \left( S(t=v)  - S(s=v) \right)
\]

In contrast to the Quantum version of the classical $A^\ast$-algorithm \cite{bauckhage2018towards}, the presented optimization function does not rely on classical preprocessing. In our solution, the number of binary variables and hence the number of Qubits required by QAOA equals the total number of arcs, whereas the solution in \cite{bauckhage2018towards} is quadratic in the number of vertices.

\section{Time-dependent phenomena}
\label{sec: time dependent phenomena}

In order to account for time-dependent phenomena like forecasted traffic or weather, a different mathematical model is required. First, we modify the graph $G$ by adding arc $(d,d)$ of weight $0$. If this arc already exists, simply set its weight to zero. Also, delete all arcs with source $d$.

\begin{figure}
	\centering
	\begin{tikzpicture}
		\node at (0, 3) {$V_0$};
		\node at (1, 3) {$V_1$};
		\node at (2, 3) {$V_2$};
		\node at (3, 3) {$V_3$};
		\node at (4, 3) {$V_4$};
		\node at (5, 3) {$V_5$};
		\node[shape=circle,draw=black] (o-0) at (0,0) {o};
		\node[shape=circle,draw=black] (5-1) at (1,2) {5};
		\node[shape=circle,draw=black] (1-1) at (1,1) {1};
		\node[shape=circle,draw=black] (2-1) at (1,0) {2};
		\node[shape=circle,draw=black] (3-1) at (1,-1) {3};
		\node[shape=circle,draw=black] (1-2) at (2,2) {1};
		\node[shape=circle,draw=black] (6-2) at (2,1) {6};
		\node[shape=circle,draw=black] (d-2) at (2,0) {d};
		\node[shape=circle,draw=black] (2-2) at (2,-1) {2};
		\node[shape=circle,draw=black] (4-2) at (2,-2) {4};
		\node[shape=circle,draw=black] (8-3) at (3,2) {8};
		\node[shape=circle,draw=black] (6-3) at (3,1) {6};
		\node[shape=circle,draw=black] (d-3) at (3,0) {d};
		\node[shape=circle,draw=black] (4-3) at (3,-1) {4};
		\node[shape=circle,draw=black] (7-3) at (3,-2) {7};
		\node[shape=circle,draw=black] (8-4) at (4,1) {8};
		\node[shape=circle,draw=black] (d-4) at (4,0) {d};
		\node[shape=circle,draw=black] (7-4) at (4,-1) {7};
		\node[shape=circle,draw=black] (d-5) at (5,0) {d};
		\draw[->] (o-0) -- (5-1);
		\draw[->] (o-0) -- (1-1);
		\draw[->, red] (o-0) -- (2-1);
		\draw[->] (o-0) -- (3-1);
		\draw[->] (5-1) -- (1-2);
		\draw[->] (1-1) -- (6-2);
		\draw[->] (2-1) -- (6-2);
		\draw[->] (2-1) -- (d-2);
		\draw[->, red] (2-1) -- (4-2);
		\draw[->] (3-1) -- (4-2);
		\draw[->] (3-1) -- (2-2);
		\draw[->] (1-2) -- (6-3);
		\draw[->] (6-2) -- (8-3);
		\draw[->] (6-2) -- (d-3);
		\draw[->] (d-2) -- (d-3);
		\draw[->, red] (4-2) -- (d-3);
		\draw[->] (4-2) -- (7-3);
		\draw[->] (2-2) -- (6-3);
		\draw[->] (2-2) -- (d-3);
		\draw[->] (2-2) -- (4-3);
		\draw[->] (6-3) -- (8-4);
		\draw[->] (6-3) -- (d-4);
		\draw[->, red] (d-3) -- (d-4);
		\draw[->] (4-3) -- (d-4);
		\draw[->] (4-3) -- (7-4);
		\draw[->] (7-3) -- (d-4);
		\draw[->] (7-4) -- (d-5);
		\draw[->, red] (d-4) -- (d-5);
	\end{tikzpicture}
	\caption{The graph $G$ of figure \ref{fig:single-vehicle routing} cut into time slices and turned into a directed acyclic graph $G'$. The time slices $V_0$ and $V_{c_\text{max}}$, $c_\text{max}=5$ contain only origin and destination and are dropped in the formulation of the optimization function.}
	\label{fig: sliced graph}
\end{figure}
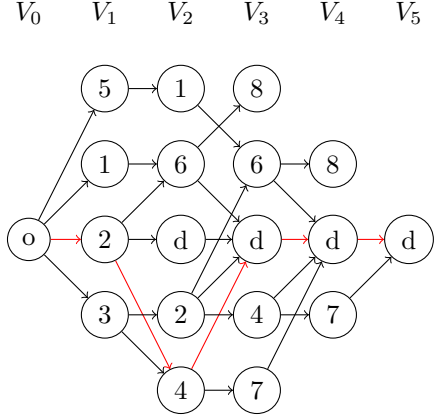

Second, we create a new directed acyclic graph $G'$ out of the modified $G$. For this, introduce time slices $c=0, ..., c_\text{max}$. The total number of slices $c_\text{max}+1$ equals the number of vertices along the longest route and is bounded by $|V|$ from above. To each time slice $c$ associate a set of vertices $V_c$. We set $V_0:=\{o\}$ and $V_1$ contains all vertices whose source equals the route origin $s=o$. Continuing iteratively, $V_{c+1}$ contains all target vertices of arcs with source in $V_c$.
\[
	V_{c+1} := \{ t \vert \text{ for all }(s,t)\in A \text{ with } s\in V_{c} \}
\]
This iterative definition terminates when $V_{c}=\{d\}$ and we set $c_\text{max} := c$. Note that the existence of cycles in the original graph $G$ may force $V_{c}\neq\{d\}$ for all $c$. In practice, one either has a good estimate for $c_\text{max}$, or one sets $c_\text{max}:=|V|-1$. Finally, we define the set of vertices $V'$ of $G'$ to be the disjoint union of the vertices in all time slices.
\[
	V':=\mathop{\dot{\bigcup}}_{c=0, ..., c_\text{max}} V_c.
\]
The arcs and arc weights of $G$ naturally carry over to build the arcs of $G'$. Namely, there is an arc from $v\in V_c$ to $v'\in V_{c+1}$ if there is an arc from $v$ to $v'$ in $G$. For a visualization of the newly created graph $G'$, see figure \ref{fig: sliced graph}.

Moving to the optimization polynomial, the binary variables $X_{(c,v)}\in\{0,1\}$ are now associated to all vertices in all time slices. In each time slice $c=0, ..., c_\text{max}$, the vehicle is required to be at exactly one vertex, which is expressed by the term
\begin{equation}\label{eq:time-dependent penalty}
	P\cdot\sum_{c=0}^{c_\text{max}}\left( \sum_{v \in V_c} X_{(c,v)} -1 \right)^2
\end{equation}
in the optimization function. The selected vertices must be connected by valid arcs whose total sum of weights is to be minimized. This gives rise to the second and final term in the polynomial:
\begin{equation}\label{eq:time costs}
	\sum_{c=0}^{c_\text{max}-1} \sum_{\substack{v\in V_c \\ v'\in V_{c + 1}}} d_{v,v'} X_{(c, v)} X_{(c+1, v')}
\end{equation}
Here, the cost factor $d_{v,v'}$ is either the weight of the arc from vertex $v$ to vertex $v'$ or the penalty value $P$.
\[
	d_{v, v'}:= \begin{cases}
		w(v,v') & \text{if such an arc exists} \\
		P & \text{otherwise}
	\end{cases}
\]
Finally, we fix the variables $X_{(0,o)} = X_{(c_\text{max}, d)}:=1$ and $X_{(c_\text{max},v)}=0$ for all $v\in V_{c_\text{max}}\setminus\{d\}$ in both terms \eqref{eq:time-dependent penalty} and \eqref{eq:time costs}.

Because of the classical preprocessing when turning the original graph $G$ into $G'$, this method might not be suited for one-time calculations of a time-dependent single-vehicle route. However, once the preprocessing has been performed, simple weight-adjustments allow for quick calculations of various scenarios. This in particular applies to aircraft-control, where routes have to be calculated every few minutes with rapidly changing exterior weather and wind conditions. We comment more on this at the end of section \ref{sec:MAC}.

\section{Travelling salesperson}
\label{sec:tsp}

Consider a salesperson supposed to travel through each of $|V|$ cities exactly once. The standard solution in the literature is the following \cite{IsingOverview}:

In the previously discussed time-dependent model, set $d=o$ and $c_\text{max} := |V|-1$. Furthermore, let $V_c :=V\setminus \{o\}$ for all $c=1, ..., c_\text{max}-1$. Finally, consider an optimization function built from the two terms of the previous section together with the following penalty term.
\[
	P\cdot \sum_{v\in V\setminus\{o\}} \left( \sum_{c=1}^{c_\text{max}-1} X_{(c,v)} -1  \right)^2
\]
This term ensures that each vertex is chosen exactly once.

Overall, the optimization function depends on $(|V|-1)^2$ variables and the optimization problem can hence be solved on quantum computers with the same number of Qubits.

However, the number of degrees of freedom in this model and hence the Qubit number is unnecessarily high: If the graph is not all-to-all connected, not all vertices can be reached from the origin and the first time slice contains vacuous vertices. The preprocessing of section \ref{sec: time dependent phenomena} cuts two cones out of the naive $(|V|-1)^2$-cube: One whose tip is the origin $o$ and one with tip $d$. The reduction in Qubit numbers is higher the lower the connectivity of the graph. Overall, $(|V|-1)^2$ is merely the upper bound for the number of binary variables.

\section{Collision-free multi-vehicle routing}
\label{sec:MAC}

We finally seek to find paths for $N$ vehicles such that no two vehicles appear at the same vertex in any point of time. For now, we assume that all vehicles travel at the same speed and each arc requires exactly one time step. However as discussed in the end, arbitrary (and even varying) speeds and stochastic external influences can naturally be incorporated.

First, for each vehicle $i=1, ..., N$ a subgraph $G_i$ of $G$ is created. The selection of this subgraph is important: The fewer vertices and arcs are selected, the fewer variables does the final optimization function depend on. Introducing time slices $c=0, ..., c_\text{max}$, each graph $G_i$ gives rise to a new graph $G_i'$ with sets of vertices $V_{(i,c)}$. Here, $c_\text{max}$ is fixed for all $i$. This time, the binary variables $X_{(i,c,v)}\in\{0,1\}$ are associated to a vehicle $i$, a time slice $c$ and a vertex $v\in V_{(i,c)}$.

For each $i$, the total optimization function contains both terms \eqref{eq:time-dependent penalty} and \eqref{eq:time costs} of section \ref{sec: time dependent phenomena} with $(c,v)$ replaced by $(i,c,v)$ and $V_c$ replaced by $V_{(i,c)}$. Hence, the minimum of this function selects shortest paths for each vehicle. In order to forbid collisions, one further term is introduced. For a given $c=0, ..., c_\text{max}$ and $v\in V$, let
\[
	D(c,v):=\sum_{i=1}^N \delta_{v\in V_{(i,c)}} X_{(i,c,v)}
\]
be the sum over all binary variables associated to $v$ which appear in the the same time slice of multiple vehicles. The additional term in the optimization function is then
\[
	P\cdot \sum_{c=1}^{c_\text{max}} \sum_{v\in V} \left( D(c,v) -1 \right) \cdot D(c,v)
\]
For $P$ big enough, this term forces maximally one variable in $D(c,v)$ to be set to $1$. In other words, no two vehicles may be at the same vertex at the same time.

The presented model can be adapted to include arbitrary vehicle speed by fixing an interval, e.g. 10 minutes, between the time slices $c$. All arcs $a\in A$ must then be traversable in an integer multiple $m_a$ of the interval length. Subsequently, $G$ is modified by decomposing $a$ into $m_a$ arcs connecting $m_a-1$ new vertices in line. Thus, all arcs in $G$ are again traversable within one time-step. This way, one can also consider multiple aircrafts with different speeds and speeds varying over time.

By construction, the presented model is most efficient when many vehicles have to take similar routes and the subgraphs $G_i$ and their time slices do not have to be recalculated each time. A natural field of application are therefore aviation industries and multi-aircraft-control. Because aircrafts generally follow rather rigid paths, the reduction from the total graph $G$ to the subgraphs $G_i$ reduces the problem size dramatically.

Similar to the seminal paper \cite{doi:10.1002/atr.5670260205}, the presented model is purely deterministic. In particular, selected routes may vary drastically upon small changes of the input parameters. In order to circumvent this dependence, arc weights can be assigned probabilities. Future weather conditions can then be incorporated with the help of Markov chains \cite{1696410, 1384440}. Research in this directed is relegated to future work.

\section*{Acknowledgements}

This work is supported by the German national initiative PlanQK.

\appendix

\section{Review of Quantum optimization methods}

\subsection{Adiabatic Quantum Computing}

Quantum computation by adiabatic evolution as proposed in \cite{farhi2000quantum} is an optimization algorithm running on dedicated Quantum hardware. The basic idea is the following.

Let $C(X_1, ..., X_B)$ be a quadratic optimization function in $B$ binary variables $X_i\in\{0,1\}$. Construct the problem Hamiltonian $\mathcal{H}_P$
\[
	\mathcal{H}_P \ket{X_0}...\ket{X_B} = C(X_1, ..., X_B) \ket{X_0}...\ket{X_B}.
\]
Prepare your Quantum system in an easy-to-construct ground state $\ket 0$ of a simple initial Hamiltonian $\mathcal{H}_I$. Finally, let the system evolve in time from $t=0$ to $t=T$ along some monotonic curve $s(t)\in[0,1], s(0)=0, s(T)=1,$ according to the Hamiltonian
\[
	\mathcal{H}(t) := \mathcal{H}_I\cdot (1-s(t)) + \mathcal{H}_P \cdot s(t).
\]
After the evolution, the system is in the state $U(T)\ket 0$ where the time-evolution operator $U$ is the solution to the Schroedinger equation with respect to $\mathcal H (t)$.

If the energy gap between ground state and first exited state is greater than zero throughout the evolution and $T$ is chosen large enough, $U(T)\ket 0$ is the ground state of $\mathcal{H}_P$ at $t=T$ by the adiabatic theorem.

\subsection{QAOA}

The Quantum Approximate Optimization Algorithm \cite{farhi2014quantum} is a hybrid quantum-classical algorithm approximating the adiabatic evolution on gate-model Quantum hardware. Its concept can be summarized as follows.

Trotterizing the time-evolution operator $U$ of adiabatic Quantum computing into $p$ steps gives
\begin{align*}
	U(T) &\approx \prod_{k=1}^p e^{-\frac i \hbar \cdot \mathcal{H}(k\cdot\delta t)\cdot \delta t} \\
	&\approx e^{-i \beta_p \mathcal{H}_I} e^{-i \gamma_p \mathcal{H}_P} ... e^{-i \beta_1 \mathcal{H}_I} e^{-i \gamma_1 \mathcal{H}_P}.
\end{align*}
In the second step, we have linearized the exponential by suppressing higher order commutators in the Baker-Campbell-Hausdorff formula. The parameters $\beta_k$ and $\gamma_k$ depend on the form of $s(t)$.

The individual factors in the trotterized form of $U(T)$ can easily be implemented as gates on a universal quantum computer. For fixed $p$, $\beta_k$ and $\gamma_k$, QAOA evaluates the trotterized version of
\begin{equation}\label{eq:QAOA evaluation}
	\left(\bra 0  U(T)^\dagger \right) \hat C \left(U(T)\ket 0 \right)
\end{equation}
on a quantum computer. A classical optimization algorithm (e.g. gradient descent) now varies $\beta_k, \gamma_k$ while treating $p$ as a fixed hyperparameter. For every set of parameters the quantum computer evaluates \eqref{eq:QAOA evaluation} until the classical algorithm terminates. A Quantum measurement of $U(T)\ket 0$ for the final parameters $\beta_k, \gamma_k$ reveals the state minimizing the optimization function $C$.

\newpage

\bibliographystyle{unsrt}
\bibliography{references}

\begin{thebibliography}{10}

\bibitem{farhi2014quantum}
Edward Farhi, Jeffrey Goldstone, and Sam Gutmann.
\newblock A quantum approximate optimization algorithm, 2014.

\bibitem{farhi2000quantum}
Edward Farhi, Jeffrey Goldstone, Sam Gutmann, and Michael Sipser.
\newblock Quantum computation by adiabatic evolution.
\newblock {\em arXiv preprint quant-ph/0001106}, 2000.

\bibitem{farhi2002quantum}
Edward Farhi, Jeffrey Goldstone, and Sam Gutmann.
\newblock Quantum adiabatic evolution algorithms versus simulated annealing.
\newblock {\em arXiv preprint quant-ph/0201031}, 2002.

\bibitem{crosson2014different}
Elizabeth Crosson, Edward Farhi, Cedric Yen-Yu Lin, Han-Hsuan Lin, and Peter
  Shor.
\newblock Different strategies for optimization using the quantum adiabatic
  algorithm.
\newblock {\em arXiv preprint arXiv:1401.7320}, 2014.

\bibitem{Preskill_2018}
John Preskill.
\newblock Quantum computing in the nisq era and beyond.
\newblock {\em Quantum}, 2:79, Aug 2018.

\bibitem{farhi2017quantum}
E.~Farhi, J.~Goldstone, S.~Gutmann, and H.~Neven.
\newblock Quantum algorithms for fixed qubit architectures, 2017.

\bibitem{bauckhage2018towards}
C~Bauckhage, E~Brito, K~Cvejoski, C~Ojeda, J~Sch{\"u}cker, and R~Sifa.
\newblock Towards shortest paths via adiabatic quantum computing.
\newblock 2018.

\bibitem{IsingOverview}
Andrew Lucas.
\newblock Ising formulations of many np problems.
\newblock {\em Frontiers in Physics}, 2:5, 2014.

\bibitem{doi:10.1002/atr.5670260205}
Lucio Bianco and Maurizio Bielli.
\newblock Air traffic management: Optimization models and algorithms.
\newblock {\em Journal of Advanced Transportation}, 26(2):131--167, 1992.

\bibitem{1696410}
A.~{d'Aspremont}, D.~{Sohier}, A.~{Nilim}, L.~{El Ghaoui}, and {Vu Duong}.
\newblock Optimal path planning for air traffic flow management under
  stochastic weather and capacity constraints.
\newblock In {\em 2006 International Conference on Research, Innovation and
  Vision for the Future}, pages 1--6, 2006.

\bibitem{1384440}
A.~{Nilim} and L.~{El Ghaoui}.
\newblock Algorithms for air traffic flow management under stochastic
  environments.
\newblock In {\em Proceedings of the 2004 American Control Conference},
  volume~4, pages 3429--3434 vol.4, 2004.

\end{thebibliography}

\end{document}